\documentclass[journal]{IEEEtran}
\usepackage{amsmath,amsfonts}     
\usepackage{algorithmic}           
\usepackage{array}                     
\usepackage[caption=false,font=normalsize,labelfont=sf,textfont=sf]{subfig}            
\usepackage{textcomp}        
\usepackage{stfloats}             
\usepackage{url}        
\usepackage{verbatim}        
\usepackage{graphicx}         
\usepackage{balance}     
\usepackage{enumitem}
\usepackage{booktabs}
\usepackage[table]{xcolor}                                                                          
\usepackage{cite}                                                               
\def\BibTeX{{\rm B\kern-.05em{\sc i\kern-.025em b}\kern-.08em                                       
  T\kern-.1667em\lower.7ex\hbox{E}\kern-.125emX}}                        
\begin{document}                                                                                             
\title{Enhancing BGP Security by Understanding BGP's Language with LLMs}                            
              
\author{Heng~Zhao,
      Ruoyu~Wang,
      Tianhang~Zheng,~\IEEEmembership{Member,~IEEE,}
      Qi~Li,~\IEEEmembership{Senior Member,~IEEE,}
      Bo~Lv,                 
      Yuyi~Wang,
      and~Wenliang~Du,~\IEEEmembership{Fellow,~IEEE,}%
\thanks{This work was supported by the Open Research Fund of the State Key Laboratory of Internet Architecture, Tsinghua University, Beijing 100084, China. \textit{(Corresponding author: Tianhang Zheng.)}}%
\thanks{Heng Zhao is with The State Key Laboratory of Blockchain and Data Security, Zhejiang University, Hangzhou, Zhejiang, China, and also with The State Key Laboratory of Internet Architecture, Tsinghua University, Beijing 100084, China (e-mail: hengzhao02@zju.edu.cn).}%
\thanks{Ruoyu Wang is with The State Key Laboratory of Blockchain and Data Security, Zhejiang University, Hangzhou, Zhejiang, China, and also with The University of Hong Kong, Hong Kong, China (e-mail: u3645257@connect.hku.hk).}%
\thanks{Tianhang Zheng is with The State Key Laboratory of Blockchain and Data Security, Zhejiang University, Hangzhou, Zhejiang, China, and also with The State Key Laboratory of Internet Architecture, Tsinghua University, Beijing 100084, China (e-mail: zthzheng@zju.edu.cn).}%
\thanks{Qi Li is with The State Key Laboratory of Internet Architecture, Tsinghua University, Beijing 100084, China (e-mail: qli01@tsinghua.edu.cn).}%
\thanks{Bo Lv is with The State Key Laboratory of Blockchain and Data Security, Zhejiang University, Hangzhou, Zhejiang, China (e-mail: bolv@zju.edu.cn).}%
\thanks{Yuyi Wang is with CRRC Zhuzhou Institute \& Tengen Intelligence Institute, China (e-mail: yuyiwang920@gmail.com).}%
\thanks{Wenliang Du is with The State Key Laboratory of Blockchain and Data Security, Zhejiang University, Hangzhou, Zhejiang, China (e-mail: wenliangdu@zju.edu.cn).}%
}
\maketitle       
\begin{abstract}                                                                                    
The trust-based nature of Border Gateway Protocol (BGP) makes it vulnerable to disruptions such as prefix hijacking and misconfigurations, posing significant threats to routing stability. 
Traditional BGP anomaly detection relies on manual inspection with poor scalability and efficiency. Machine/Deep Learning (M/DL)-based approaches automate detection but suffer from suboptimal precision, limited generalizablity and high retraining cost.
This is because existing M/DL-based methods mainly focus on capturing topological structures rather than comprehensive semantic characteristics of Autonomous Systems (ASes), which may assign dissimilar embeddings for functionally similar but topologically distant ASes or generate embeddings that are highly sensitive to topology evolution.
To address these limitations, we propose BGPShield, a novel anomaly detection framework built on an Adaptive LLM BGP Encoder, which can capture the \emph{Behavior Portrait} and \emph{Routing Policy Rationale} of each AS beyond the topology information, such as operational scale and global role in the Internet.
Inspired by multimodal LLMs (MLLMs), we propose an adaptive LLM BGP encoder to generate embedding for representing both routing behaviors and semantics of ASes by training the encoder on a contrastive loss. 
Furthermore, we introduce SAM-ED algorithm to quantity BGP-specific semantic deviations between historical and updated paths, rather than simply accumulating distances between path elements without awareness of BGP-specific structures.
Evaluated on 16 real-world datasets, BGPShield detects 100\% of verified anomalies with an average false discovery rate below 5\%. Note that the open-source LLMs employed by BGPShield were released prior to several events used for evaluation, verifying BGPShield's generalizability on unseen events.
Furthermore, BGPShield can construct the representation for a previously unseen AS within one second, significantly outperforming BEAM which demands thorough retraining (averagely 65 hours).
\end{abstract}
 
\section{Introduction}
\label{sec:intro}
\IEEEPARstart{T}{he} Border Gateway Protocol (BGP) serves as the "GPS" of the Internet, coordinating how data travels across vast interconnected networks known as Autonomous Systems (ASes)~\cite{rfc4271}. This system relies on trust, meaning a single incorrect BGP update\footnote{BGP updates are routing messages exchanged between ASes to advertise or withdraw reachability information.} can misdirect global traffic.
Numerous incidents (\emph{e.g.,} the 2014 large-scale route leak~\cite{cisco2014} and 2021 Facebook outage~\cite{facebook2021}) underscored potential disruptions or economic losses caused by misconfigurations, malicious attacks, \emph{etc}. 
To mitigate those threats, various rule-based detection paradigms (\emph{e.g.,} RPKI validation~\cite{rfc6811} and S-BGP~\cite{839934}) have been proposed to verify route authenticity and enforce static policy checks. However, these methods typically lack adaptability to evolving attack strategies and might induce high operational overhead under complex  conditions~\cite{chen2022rov}. 
Considering the limitations of traditional methods, the community pivoted Machine Learning (ML) based mechanism as an alternative. 
Related studies~\cite{cheng2016ms, testart2019profiling,cheng2018multi,dong2021isp,al2015detecting,al2012machine,9754706,10.1145/3359992.3366640,6848023,6795963,9761992} have demonstrated that ML methods can automatically identify deviations from legitimate routing behaviors,
improving detection efficiency and precision of BGP anomaly detection. 
For instance, BEAM~\cite{298068} pioneers the integration of Deep Learning (DL) with relationship-based AS role definitions, achieving state-of-the-art (SOTA) performance with DL embeddings\footnote{In this paper, \emph{embedding} and \emph{representation} are used interchangeably, referring to the learned vectorized expression of an AS.}.
More recently, Ares~\cite{ares} and DFOH~\cite{dfoh} introduced detection paradigms based on routing tree clustering and multi-vantage point inference, respectively. 
Although statistical and M/DL-based methods have demonstrated notable progress, they cannot comprehensively capture the behavioral semantics\footnote{Behavioral semantics encompasses an AS's \textbf{routing behaviors} (how and with whom it exchanges routing information) and the \textbf{routing semantics} (\emph{e.g.,} economic rationales) that drive those tendencies.} of ASes due to inherent limitations in their architectures. 
Specifically, statistical methods compare handcrafted routing features with predefined thresholds, only relying on empirical signals without interpreting their semantic information. 
Existing M/DL based methods learn representations automatically, but they remain ineffective at capturing ASes' semantic information.
For instance, graph learning-based methods such as BEAM~\cite{298068} focus on optimizing AS representations to learn topological proximity and hierarchy within AS graphs. \emph{Thus, the resulting embedding space is governed by the ASes' topological structures rather than behavioral semantics: ASes with similar topological structures are embedded closely regardless of whether their routing behaviors or policy rationales are actually similar.} Sequential models~\cite{cheng2018multi, dong2021isp} learn temporal regularities in BGP update streams but have no mechanism to reason about the intents behind observed routing changes. 
Both limitations share a common root cause: regardless of their architectural complexity, existing methods extract structural patterns or statistical regularities from AS-level routing data, rather than the intents behind routing behaviors. 
This gives rise to two challenges for existing M/DL-based approaches:

\noindent \textbf{Detection Precision.}
Without behavioral semantics, M/DL-based AS embeddings are governed by topological structures.
This causes topologically similar but functionally different ASes to share similar embeddings, directly degrading anomaly detection precision. For instance, even BEAM cannot detect several anomalies (\emph{e.g.,} the CelerBridge event~\cite{CelerBridge2022}), highlighting the potential to further improve detection precision and false discovery rate (FDR)~\cite{fdr}.

\noindent \textbf{Generalizability for Evolving Network Conditions.}
Another limitation lies in the static nature of topology-based embeddings, which capture structural features but neglect behavioral semantic information. Thus, adapting to evolving business relationships or newly emerging ASes necessitates exhaustive retraining. Even SOTA models require periodic retraining to adapt new BGP data, limiting the real-time adaptability and generalizability in dynamic networks. 

\emph{Given previous challenges, does there exist a model architecture capable of capturing the behavioral semantics from raw AS-related data, including routing behavior tendencies and underlying policy rationales, rather than merely structural patterns or statistical regularities?
}

To address this question, we propose BGPShield, a novel anomaly detection framework built on an \textbf{A}daptive \textbf{L}LM \textbf{B}GP \textbf{E}ncoder (ALBE) that captures the behavioral semantics (details in Section~\ref{sec:profile-construction}) from all raw AS-related data. 
Notably, the primary challenge in capturing the behavioral semantics lies in extracting routing semantics. Because routing semantics, such as underlying policy rationales, are typically documented in BGP reports rather than explicitly encoded in routing or topology data, they remain inaccessible to existing methods. Pre-trained LLMs provide a solution to this challenge. Their training on large-scale corpora encompassing BGP documentation allows them to infer policy rationales from raw data via natural language understanding, whereas traditional M/DL-based methods cannot infer intents beyond structural patterns.

Consequently, inspired by multimodal LLMs (MLLMs)\cite{liu2023visualinstructiontuning}, we propose ALBE, a unified framework that integrates the semantic reasoning of LLMs with a lightweight adapter.
By training this architecture with a contrastive loss, \textbf{ALBE can embed AS profiles into BGP embeddings whose distances represent behavioral similarities beyond mere topology.}
This design ensures that ALBE and its generated embeddings remain effective across BGP evolution, supporting embedding generation for a previously unseen AS within one second without retraining.
Based on AS embeddings, we introduce SAM-ED (\textbf{S}emantic \textbf{A}nomaly-\textbf{M}aximized \textbf{E}lastic \textbf{D}istance), a BGP-specific alignment algorithm that \textbf{quantifies the semantic deviation between historical and updated paths in the BGP domain space}, rather than merely accumulating distances between path elements without awareness of BGP-specific structures.
Specifically, SAM-ED incorporates two BGP-aware designs: an Anomaly-prioritized AS Set Resolution preserving anomalous deviations introduced by AS sets, and an Explainable Alignment Traceback pinpointing the exact positions of semantic behavioral divergence along the aligned path (see Section~\ref{sec:path-diff}).
Anomalous route changes are thereby identified by our BGP Anomaly Detector, as legitimate route changes typically exhibit low deviation from historical paths in the BGP domain space (see Section~\ref{sec:path-diff-evaluation}).
Finally, BGPShield aggregates individual anomalous BGP updates 
into events based on affected prefixes and responsible ASes. 

We further evaluate BGPShield on 16 real-world datasets. \emph{Among them, several datasets were collected after the public release of the open-source LLMs employed by BGPShield, verifying the generalizability of BGPShield on unseen events.} (detailed in Section~\ref{sec:dataset}).
The experimental results show that \textbf{BGPShield detects 100\% of publicly reported BGP anomaly events}, 
while the SOTA method BEAM only identifies 81\% of them.
Simultaneously, BGPShield achieves an average FDR that is 2-3$\times$ lower than BEAM.
\textbf{Even with 25\% noise in AS-level information, BGPShield still sustains 98.8\% precision}, indicating robust generalizability to evolving BGP networks.
Moreover, BGPShield can generate embeddings for a newly observed AS within one second, whereas BEAM requires thorough retraining (averaging 65 hours) on the entire AS graph. This capability enables BGPShield to seamlessly adapt to network evolution while maintaining long-term (over 5 years) effectiveness without retraining.

Our contributions can be summarized as follows:
\begin{itemize}[itemsep=0em, align=parleft, left=0pt]

    \item We design the \textbf{ALBE, an Adaptive LLM BGP Encoder} trained on a contrastive loss to embed AS profiles into the BGP embeddings. ALBE can semantically capture the routing behaviors and routing semantics beyond topology, supporting embedding generation for a previously unseen AS within one second without retraining.
    \item We propose \textbf{SAM-ED} algorithm, a BGP-specific alignment algorithm that quantifies semantic deviations between historical and updated paths with Anomaly-prioritized AS Set Resolution and Explainable Alignment Traceback, rather than merely accumulating distances between path elements without awareness of BGP-specific structures.
    \item We implement and evaluate BGPShield\footnote{Source code:  \url{https://github.com/ZJU-LLM-Safety/BGPShield}} 
    on 16 real-world datasets (several were collected after the release of the LLMs employed by BGPShield), achieving \textbf{100\% detection of verified anomalies with an average FDR 2–3$\times$ lower than the SOTA method (BEAM)} and maintaining \textbf{98.8\% precision under 25\% AS-level information noise}, demonstrating high detection precision and robust generalizability to evolving BGP networks and unseen events.
\end{itemize}

\section{Related Work}
\label{sec:rela-work}


Traditional anomaly detection methods can be classified into three categories by their data sources: 
\textbf{control-plane} methods~\cite{10.5555/1267336.1267347, 8481500, testart2019profiling, 10.1145/1096536.1096542} maintain mappings of prefixes to ASes under normal circumstances, which are typically obtained from globally distributed route collectors, and compare each new update against this reference to detect anomalies.
\textbf{data-plane} methods~\cite{7778593, 6459962} alarm anomalies by actively probing real-time data obtained from multiple sensors and analyzing the reachability from certain hosts to the target prefix.  
\textbf{hybrid} methods~\cite{4223210, 7460217, DBLP:conf/ndss/VervierTD15, 10.1145/2398776.2398779} combine both control‑plane and data‑plane information to analyze suspicious updates, aiming to reduce false positives by corroborating anomalies across multiple observation layers. 
While these traditional approaches have advanced the understanding of BGP anomalies, these approaches rely on static configurations and manual correlation, limiting scalability and responsiveness to dynamic BGP networks.

To overcome these limitations, researchers have turned to M/DL for automated anomaly detection~\cite{cheng2016ms, testart2019profiling, cheng2018multi, dong2021isp, al2015detecting, al2012machine, 9754706, 10.1145/3359992.3366640, 6848023, 6795963, 9761992}.
Cheng \emph{et al.}~\cite{cheng2018multi} 
introduced a multi-scale LSTM model that incorporates discrete wavelet transforms and hierarchical attention mechanisms 
to extract temporal features for route leak detection, though its applicability remains restricted to specific anomaly types.
Dong \emph{et al.}~\cite{dong2021isp} proposed a weakly supervised detection system using knowledge distillation from third-party detectors and self-attention LSTMs~\cite{6795963} to address label scarcity and noise. 
Shapira \emph{et al.}~\cite{9754706} 
developed AP2Vec based on BGP2Vec~\cite{9761992} to embed AS numbers and IP prefixes, enabling a two-phase detection process that identifies structural changes without requiring labeled data. 
Although AP2Vec improved scalability and practicality, it exhibited limited adaptability to novel attack patterns and incurred high retraining costs.
Notably, Chen \emph{et al.}~\cite{298068} proposed BEAM, a representation learning model that builds AS-level business relationship graphs and constructs embeddings to encode both topological proximity and hierarchy, 
achieving SOTA performance without the need for manual feature engineering.
However, its precision and generalizability remain limited, as the embeddings primarily reflect structural topology rather than the comprehensive behavioral semantics of ASes.
More recently, Ares~\cite{ares} evaluates routing tree evolutions via weighted tree edit distance and clustering to identify hijackings. 
Although Ares enables high-precision detection without labeled data, its reliance on specific parameter configurations may result in performance variations across diverse anomaly incidents.
Given Ares's algorithmic sensitivities (further discussed in Section~\ref{sec:experiment}), and considering that DFOH targets forged-origin hijacks rather than generalized BGP anomalies, 

\section{Preliminary Data Preparation}
\label{sec:dataset}

This section describes our Route Monitor, a system designed to acquire and process raw BGP routing data, yielding high-quality route changes suitable for anomaly detection. 
By integrating Routing Information Base (RIB) snapshots and UPDATE files from RouteViews~\cite{routeviews}, the Monitor establishes a unified data processing pipeline that provides a consistent foundation for downstream detection tasks.

Specifically, route change construction establishes a baseline using the most recent RIB snapshot (a global view of prefix-to-AS path mappings from about 40 vantage points). \textbf{Notably, this baseline does not require absolute purity because Route Monitor records the divergence between the baseline and the updates. Any behavioral shift, whether transitioning into an anomalous state or recovering from one, will be successfully captured for subsequent anomaly detection.}
For each vantage point, the mappings are indexed into prefix trees following the Longest Prefix Match (LPM) rule.
Utilizing these trees, the Monitor processes UPDATE files every 15 minutes by querying the most specific historical entry for each update.
A route change is recorded if the updated AS path diverges from the reference. For example, comparing an updated prefix \verb|*.*.153.0/24| (\verb|ASa|$\rightarrow$\verb|ASb|$\rightarrow$\verb|ASc|) with its corresponding reference \verb|*.*.152.0/22| (\verb|ASa|$\rightarrow$\verb|ASd|$\rightarrow$\verb|ASc|) yields a route change.
These differences reflect behavioral shifts but do not signify anomalies, as legitimate practices (\emph{e.g.,} traffic engineering or policy adjustments) also induce variations. BGPShield filters out anomalous route changes by detecting subtle behavioral semantic shifts along routing paths.

To comprehensively evaluate BGPShield, 
we construct 16 real-world datasets (totaling 1,327,756,266 route changes) corresponding to publicly reported anomaly incidents spanning from 2008 to 2025. Each dataset includes documented routing anomalies with ground-truth information manually derived from BGPmon~\cite{bgpmon2025}, Dyn~\cite{dyn2025}, MANRS~\cite{manrs2025}, and APNIC~\cite{apnic2025}.
The 16 datasets comprise 6 \emph{Origin Change}, 5 \emph{Path Manipulation} and 5 \emph{Route Leak}, comprehensively covering common anomaly types. 
To verify the generalizability of BGPShield, we include 3 incidents that occurred after the latest public release (May 29, 2025) of the LLMs employed by BGPShield, 
thereby ensuring zero prior exposure during pre-training.
For each dataset, we extracted 24-hour UPDATE windows (12 hours before and after the reported incident) from RouteViews to capture both anomalous and sufficient legitimate routing dynamics. 
\textbf{Notably, for real-time detection, the Route Monitor maintains a persistent baseline to intercept anomalies upon arrival, preventing ongoing attacks from appearing in both the baseline and updates simultaneously.}
\section{BGPShield System}
\label{sec:bgpshield}

\begin{figure*}[t]
    \centering  \includegraphics[width=\linewidth]{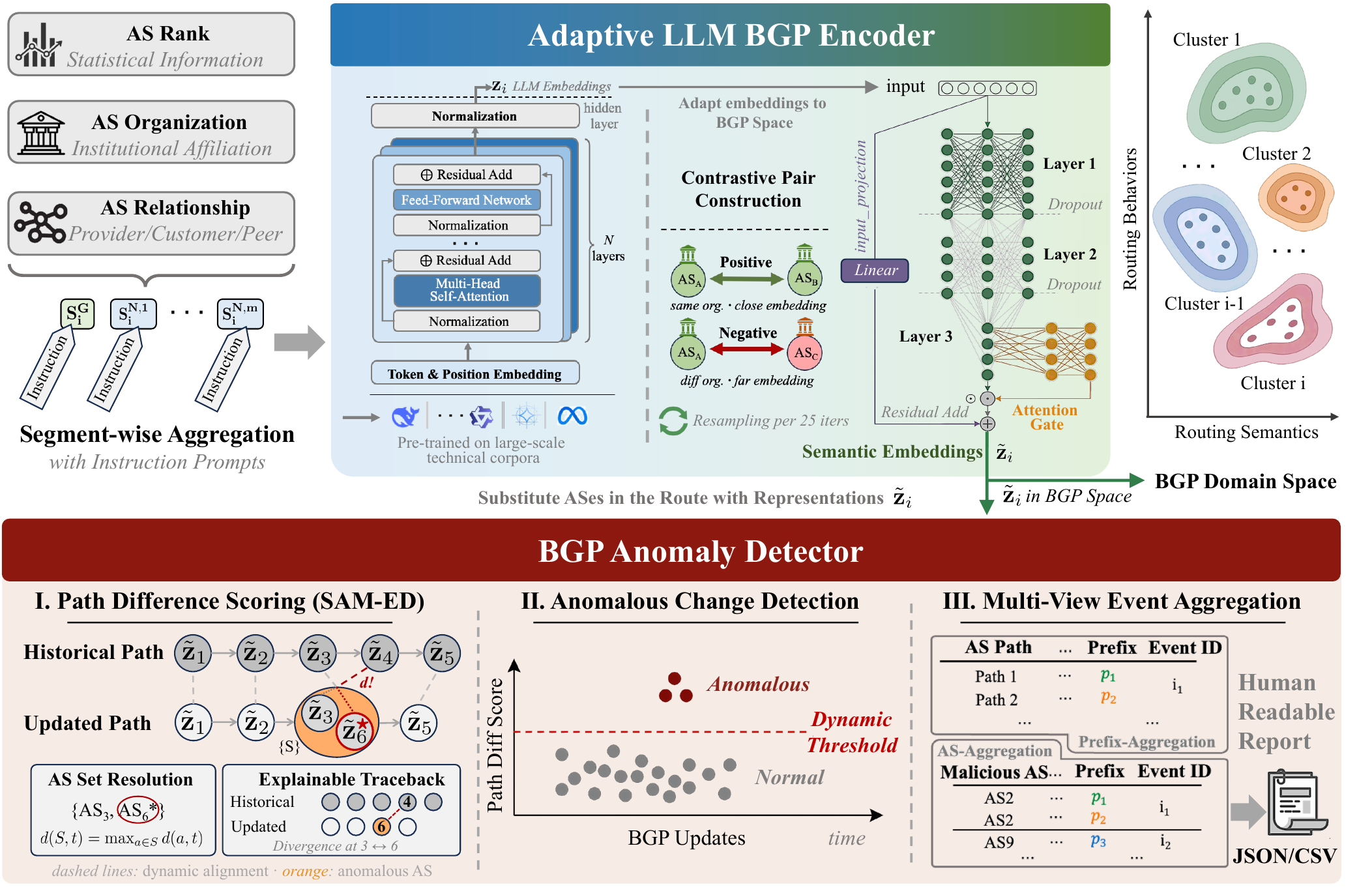}
    \caption{\label{fig:bgpshield}\textbf{BGPShield System Overview.}}
\end{figure*}

\subsection{System Overview}
\label{sec:overview}

As illustrated in Figure~\ref{fig:bgpshield}, BGPShield consists of two core modules: the Adaptive LLM BGP Encoder (ALBE) and the BGP Anomaly Detector (BAD). 
This architecture originates from the intuition that genuine routing anomalies manifest as behavioral deviations rather than mere BGP topological structures. By shifting the detection paradigm from topological structures toward behavioral semantics reasoning, BGPShield overcomes the inherent brittleness of existing statistical and M/DL-based methods which rely on topological structures or statistical regularities.

Inspired by recent advances in MLLMs, 
ALBE serves as the semantic engine of BGPShield,
integrating the semantic reasoning of an LLM with a lightweight adapter into a unified architecture.
Trained on a contrastive loss, ALBE takes AS profiles formulated by a segment-wise aggregation scheme with instruction prompts as input, and outputs BGP semantic embeddings, capturing the routing behaviors and routing semantics beyond topology. 
Subsequently, BAD utilizes these semantic embeddings to identify anomalies through the SAM-ED algorithm, which dynamically align the embedding sequences of historical and updated paths.
This process evaluates behavioral inconsistencies by computing a cumulative path difference score. BAD identifies anomalous route changes by applying an adaptive threshold derived from statistical distributions of these scores. Finally, a multi-view aggregation scheme consolidates individual anomalies into unified events, pinpointing the responsible ASes and affected prefixes. 

\subsection{AS Semantic Embedding Generation}
\label{sec:embedding-gen}

\subsubsection{Segment-wise AS Profile Construction}
\label{sec:profile-construction}

To construct AS profiles that expose behavioral semantics to the LLM, we organize AS attributes around two features:

\noindent\textbf{Feature A (\emph{Behavioral Portrait})} captures operational behaviors (\emph{e.g.,} topology and operational scale), reflecting \emph{how} an AS interacts and exchanges routes with others in the global routing ecosystem, distinguishing functional roles that purely topological structures cannot capture.

\noindent\textbf{Feature B (\emph{Routing Policy Rationale})} 
reflects \emph{why} an AS adopts specific routing behaviors, covering underlying policy rationales (\emph{e.g.,} economic motivations and strategic reasoning). These rationales remain relatively stable even as business relationships evolve, supporting robust detection performance across long-term network changes.

To cover both features, we organizes heterogeneous BGP metadata into two categories: temporal stability (long-term stable \emph{vs.} dynamic) and information granularity ( statistics \emph{vs.} per-neighbor records), forming the structured AS profile $\mathcal{S}_i$ for each AS $i$ (details in Supplemental materials).

\textbf{Global Identity Attributes.}
This category integrates relatively stable attributes to establish the fundamental identity and general routing role of an AS, including institutional information such as the AS number and affiliated organization. Beyond basic identity, aggregated connectivity indicators, including customer cone size (the set of ASes reachable through an AS's downstream customers), and advertised prefix counts \emph{etc.}, quantify the BGP routing service coverage of an AS. 
These attributes persist at a stable order of magnitude as BGP networks evolve, primarily supporting \textbf{Feature A} by exposing operational scale differences that local topology alone cannot reveal. For example, two ASes may share comparable immediate neighbor counts, yet differ substantially in customer cone size and prefix coverage, indicating their distinct functional roles in the global routing ecosystem.
The aggregate distribution of neighbors further offers secondary signals for \textbf{Feature B}, reflecting the AS's general commercial positioning. For instance, an AS with a high customer-to-provider ratio typically functions as a transit provider motivated by downstream revenue, while one dominated by peer relationships reflects a cost-minimizing settlement-free interconnection
preference. The robustness of these attributes against network variation is further validated in Section~\ref{sec:robust-analysis}.

\textbf{Relational Context Attributes.}
Complementing the global attributes, this category formalizes
per-neighbor information to expose finer-grained
behavioral details. For each neighboring AS, we integrate its
business relationship type (\emph{i.e.,} provider, peer, or
customer) together with its own connectivity statistics. 
As the business relationship type is a direct expression of economic and strategic intent rather than topological proximity, 
these relational signals primarily support \textbf{Feature B}, reflecting the policy-level logic that governs AS graph evolution.
Beyond policy rationales, these records provide secondary support for \textbf{Feature A} by situating the focal AS within the global routing hierarchy: an AS whose customers are predominantly small single-homed ASes likely operates as a transit provider at the network periphery, whereas one whose providers are large ASes with extensive customer cones likely sits near the network edge, relying on upstream ASes for global reachability. This distinction remains invisible in statistical information but becomes discernible once the scale of each business neighbor is examined alongside the relationship type.

With the structured profile $\mathcal{S}_i$, we propose a \textbf{segment-wise aggregation} scheme with instruction prompts to encode it (details in our source code). Since the finite context window $T$ of Transformer models prevents processing thousands of neighbors in a single pass, $\mathcal{S}_i$ is partitioned into a persistent segment $\mathcal{S}_i^G$ (global attributes) and a variable segment $\mathcal{S}_i^N$ (relational attributes). $\mathcal{S}_i^N$ is further divided into $m$ disjoint subsets, forming bounded sequences $\mathcal{S}_i^j = \mathcal{S}_i^G \cup \mathcal{S}_i^{N,j}$ with $|\mathcal{S}_i^j| \le T$. 
Each sequence is paired with an instruction prompt, ready to be processed by ALBE in the subsequent stage.




\subsubsection{Adaptive LLM BGP Encoder}
\label{sec:adapter}

Behavioral semantics of ASes are typically documented in technical documents in natural language form, rather than explicitly encoded in routing or topological data. A pre-trained LLM, trained on large-scale technical corpora, is therefore well-positioned to parse routing-revelant information from AS profiles expressed in natural language, going beyond topological structures and statistical regularities that graph-based and statistical methods are confined to. Specifically, each bounded sequence $\mathcal{S}_i^j$ is independently processed by a pre-trained LLM $f(\cdot)$, and the results are averaged to produce the AS embedding:

\begin{equation}
\mathbf{z}_i = \frac{1}{m} \sum_{j=1}^{m} f(\mathcal{S}_i^j).
\end{equation}

Incorporating $\mathcal{S}_i^G$ in every pass maintains a stable identity anchor for the AS and absorbs granular behavioral details from each neighbor subset, keeping $\mathbf{z}_i$ invariant to minor fluctuations in the local adjacency set. This inductive design allows ALBE to generate embeddings for newly observed ASes without retraining, bypassing the cold-start limitations of prior DL-based methods.

Inspired by the adapter-based design in MLLMs, we further introduce the \textbf{BGP Adapter} $g_\phi: \mathbb{R}^d \to \mathbb{R}^{d'}$, a lightweight network that adapts the LLM's rich but BGP-uncalibrated embeddings into the BGP domain space, where distances reflect AS behavioral semantic similarity across the two dimensions defined above, rather than merely topological structures and statistical regularities.

The BGP Adapter is trained with a contrastive loss based on BGP domain knowledge. ASes under the same organizational identity are administered by the same entity, and therefore tend to exhibit similar routing policies and business incentives. We construct positive pairs as ASes sharing the same or similar organizational identities whose embedding distances also fall within the 25th percentile ($Q_{25}$) of the global distance distribution; negative pairs are drawn from different organizations with distances exceeding the 75th percentile ($Q_{75}$). The global distance distribution is estimated by randomly sampling AS pairs prior to training, and distance thresholds filter out ambiguous pairs near the boundary. A periodic resampling mechanism refreshes training pairs every 25 iterations based on the current embedding distribution to prevent overfitting.

Let $\tilde{\mathbf{z}}_i = g_\phi(\mathbf{z}_i)$ denote the adapted embedding for AS $i$. For a mini-batch sampled from the positive set $\mathcal{P}$ and negative set $\mathcal{N}$, the contrastive loss is defined as:

\begin{equation}
\label{math:adapter-loss}
    \mathcal{L}_{\mathrm{Adapter}} 
    = \ln(1 + \exp(\mathcal{L}_{\mathrm{pos}} - \mathcal{L}_{\mathrm{neg}})).
\end{equation}

where $\mathcal{L}_{\mathrm{pos}}$ and $\mathcal{L}_{\mathrm{neg}}$ represent the average distances for positive pairs $(\tilde{\mathbf{z}}_{i}, \tilde{\mathbf{z}}_{j_p}) \in \mathcal{P}$ and negative pairs $(\tilde{\mathbf{z}}_{i}, \tilde{\mathbf{z}}_{k_n}) \in \mathcal{N}$:

\begin{equation}
    \begin{aligned}
        &\mathcal{L}_{\mathrm{pos}} = \frac{1}{P} \sum_{p=1}^{P} 
        \frac{1}{d'} \left\lVert \tilde{\mathbf{z}}_{i} - 
        \tilde{\mathbf{z}}_{j_p} \right\rVert_2^2, \\
        &\mathcal{L}_{\mathrm{neg}} = \frac{1}{N} \sum_{n=1}^{N} 
        \frac{1}{d'} \left\lVert \tilde{\mathbf{z}}_{i} - 
        \tilde{\mathbf{z}}_{k_n} \right\rVert_2^2,
    \end{aligned}
\end{equation}



The training objective is:

\begin{equation}
\label{math:training-objective}
\min_{\phi} \, \mathbf{E}_{(i,j_p,k_n) \sim \mathcal{P} \cup \mathcal{N}} 
\left[ \mathcal{L}_{\mathrm{Adapter}}
(\tilde{\mathbf{z}}_i, \tilde{\mathbf{z}}_{j_p}, \tilde{\mathbf{z}}_{k_n}) 
\right].
\end{equation}

Since the contrastive loss is grounded in AS attributes that remain relatively stable across network evolution, ALBE learns to adapt embeddings based on behavioral patterns that persist beyond topological structures. 
The trained ALBE therefore remains effective across BGP topology evolution without retraining. Its effectiveness is further validated through ablation studies in 
Section~\ref{sec:overall-performance}.

\subsection{BGP Anomaly Detector}
\label{sec:BAD}

After generating AS representations, the \textbf{BAD} forms the core analytical module responsible for identifying, aggregating, and attributing anomalies arising from ambiguous route changes through a three-stage pipeline: The first stage computes difference scores between updated and historical routes, quantifying the magnitude of routing changes in a semantically meaningful way. The second stage identifies potential anomalies using a sliding window combined with dynamic thresholds. The final stage aggregates scattered anomalous route changes into distinct, temporally bounded anomaly events, which are then compiled into reports detailing their associated prefixes and responsible ASes.

\subsubsection{Path Difference Scoring}
\label{sec:path-diff}

Given a route change (see Section~\ref{sec:dataset}), BAD evaluates the semantic deviation between the historical and updated route by computing a path difference score. Route paths present two challenges that traditional sequence comparison techniques, such as edit distance, cannot address: path vary in length across route changes, requiring flexible alignment; and AS-level deviations must be measured in terms of routing behaviors and routing semantics rather than topological proximity alone.
Dynamic Time Warping algorithm~\cite{10.5555/3000850.3000887} provides flexible alignment of variable-length sequences, yet its distance metric operates without any awareness of BGP-specific structures such as AS sets.
To address these limitations, we introduce \textbf{SAM-ED}, an flexible alignment algorithm that measures semantic deviations along dynamically aligned AS paths, where embedding distances reflect both topological proximity and behavioral semantic similarity.
Its BGP-specific designs amplify anomalous signals and remains robust to routine routing variations.
Formally, let $\mathbf{s} = [s_1, s_2, \dots, s_n]$ and $\mathbf{t} = [t_1, t_2, \dots, t_m]$ denote two AS paths, where $s_i$ and $t_j$ are AS identifiers. SAM-ED removes consecutive duplicate ASes to eliminate artifacts from route flapping. The cleaned sequences are denoted by $\mathbf{s}_{\mathrm{c}}$ and $\mathbf{t}_{\mathrm{c}}$, with respective lengths $n'$ and $m'$. Each AS $v_i$ is represented as a reduced semantic embedding $\tilde{\mathbf{z}}(v_i) \in \mathbb{R}^{d'}$. The dissimilarity between nodes $s_i$ and $t_j$ is defined by the Euclidean distance:

\begin{equation}
    d(s_i,t_j)=\left\|\tilde{\mathbf{z}}(s_i)-\tilde{\mathbf{z}}(t_j)\right\|_2=\sqrt{\sum_{l=1}^{d^{\prime}}\left(\tilde{\mathbf{z}}_l(s_i)-\tilde{\mathbf{z}}_l(t_j)\right)^2}.
\end{equation}

The overall alignment cost between two paths is then computed via dynamic programming:

\begin{equation}
    \mathcal{C}[i,j] =
    d\bigl(s_{(\mathrm{c},i)}, t_{(\mathrm{c},j)}\bigr) +
    \min\left\{
        \begin{aligned}
        &\mathcal{C}[i{-}1,j], \\
        &\mathcal{C}[i,j{-}1], \\
        &\mathcal{C}[i{-}1,j{-}1]
        \end{aligned}
    \right\},
\end{equation}
where $\mathcal{C}[i,j]$ denotes the minimal accumulated cost of aligning the first $i$ elements of $\mathbf{s}_{\mathrm{c}}$ with the first $j$ elements of $\mathbf{t}_{\mathrm{c}}$. 
The final path difference score can be formulated as 
$D(\mathbf{s}_{\mathrm{c}},\mathbf{t}_{\mathrm{c}}) \;=\; \mathcal{C}[n', m']$, 
quantifying the deviation of embedding sequences between the two paths.

\emph{Furthermore, we introduce two BGP-aware designs to improve the responsiveness and interpretability of SAM-ED:}

\noindent \textbf{Anomaly-prioritized AS Set Resolution }.
This innovation addresses the semantic ambiguity introduced by AS sets, which are frequently observed in BGP. Traditional strategies compute the distance between an AS set and an AS directly using the first member in the set:
\begin{equation}
    d(S, t_j) = d(S[0], t_j),   
\end{equation}
which often obscure anomalies if AS $S[0]$ is benign. SAM-ED instead applies a maximum-distance rule:
\begin{equation}
    d(S, t_j) = \max_{a \in S} d(a, t_j),
\end{equation}
ensuring that the most semantically distant AS in the set dominates the alignment cost. This strategy makes the path difference metric more sensitive to irregularities by preserving anomalous deviations.

\noindent \textbf{Explainable Alignment Traceback}.
Traditional alignment metrics only output an aggregated scalar distance score or naively map to the first element in a set during subsequent traceback. In contrast, SAM-ED provides explicit, node-to-node traceability. Specifically, alongside computing the cumulative cost $\mathcal{C}[i,j]$, we maintain a mapping function that records the exact representative AS alignment at each step. By tracing back and restoring the participants responsible for the minimal cost path, SAM-ED yields an explainable alignment sequence. This transparent mapping explicitly reveals exactly where topological or semantic divergence occurs, helping to accurately pinpoint the root cause during subsequent anomaly attribution and localization.

By incorporating these two BGP-aware designs, SAM-ED achieves a more accurate and anomaly-sensitive measurement of BGP path differences.

\subsubsection{Anomalous Change Detection}
\label{sec:suspicious-detection}

Building upon the previously computed path difference scores, we implement a normalization step to eliminate the bias caused by differences in AS path lengths. Without normalization, difference scores often grow proportionally or superlinearly with path length, allowing deviations on longer paths to diminish the relative impact of changes on shorter paths. This not only reduces detection sensitivity for short paths but also undermines comparability across paths of different lengths. To address this, we normalize the difference score as:
\begin{align}\label{eq:norm-score}
C^\ast(\mathbf{s}, \mathbf{t}) = \frac{C(\mathbf{s}, \mathbf{t})}{\sum_{i=1}^{l_1-1} d(s_i, s_{i+1}) + \sum_{i=1}^{l_2-1} d(t_i, t_{i+1})},
\end{align} 
where $C(\mathbf{s}, \mathbf{t})$ is the path distance score between $\mathbf{s}$ and $\mathbf{t}$, and $\sum_{i=1}^{l_*-1} d$ is the cumulative embedding distance along a path. The normalization ensures that paths of different lengths are evaluated on a comparable scale, improving the validity and reliability of detection.

Then, BGPShield identifies anomalous route changes by comparing these normalized scores against a adaptive threshold.
As BGP routes exhibits substantial temporal fluctuations (\emph{e.g.,} transient link failures), a static threshold may overlook subtle anomalies during stable intervals or trigger excessive false alarms during volatile periods.
To address this, we adopt an adaptive thresholding strategy with a sliding window of length $w$ (one hour). 
The threshold $\theta_w$ for the current window is derived from the set of normalized scores $\mathcal{D}_{w-1}$ from the preceding interval.
To ensure robustness against extreme outliers, we apply a Winsorization~\cite{dixon1960simplified} step by excluding scores exceeding the 99.7th percentile. Let $\tilde{\mathcal{C}}_{(w-1)}$ denote this trimmed set. We define $\mu$ and $\sigma$ as the empirical mean and standard deviation of $\tilde{\mathcal{C}}_{(w-1)}$:
\begin{equation}
    \begin{aligned}
    \mu &= \frac{1}{|\tilde{\mathcal{C}}_{(w-1)}|} \sum_{C^* \in \tilde{\mathcal{C}}_{(w-1)}} C^* ,\\
    \sigma &= \sqrt{\frac{1}{|\tilde{\mathcal{C}}_{(w-1)}|} \sum_{C^* \in \tilde{\mathcal{C}}_{(w-1)}} (C^* - \mu)^2}.
    \end{aligned}
\end{equation}
The dynamic threshold is then calculated as $\theta_w = \mu + 4\sigma$. The factor of $4\sigma$ corresponds to a Gaussian exceedance probability of approximately $3 \times 10^{-5}$, a setting validated through empirical analysis in Section~\ref{sec:path-diff-evaluation}. By dynamically adjusting $\theta_w$, BGPShield maintains sensitivity to genuine anomalies despite routine routing variations.

\subsubsection{Multi-View Event Aggregation}
\label{sec:event-aggregation}

Path-level detection alarms numerous anomalous route changes, producing a set of route-level anomaly data. To enable event-level analysis and improve interpretability, 
BGPShield aggregates route-level alarms into interpretable events via two axes: \emph{the prefix axis} for impact scope and \emph{the AS axis} for attribution.

On \emph{the prefix axis}, anomalies affecting the same prefix are grouped and chronologically ordered. A fixed-length sliding window $w$, identical to that used in Section~\ref{sec:suspicious-detection}, counts how many vantage points report anomalies for each prefix within each interval, yielding an observation-count curve.
Moreover, we set the knee point\footnote{The knee point denotes where the curve's rate of change shifts most sharply, serving as an effective balance between sensitivity and false alarms.} of the observation-count curve as the threshold to filter transient noise.
On \emph{the AS axis}, BGPShield consolidates prefix-level incidents by identifying the ASes responsible for the anomalies. For each prefix event, BAD intersects AS sets from historical and updated routes across all deviations to pinpoint ASes, as the misbehaved ASes often impact multiple prefixes simultaneously. 
Finally, BAD links prefix events that overlap temporally and share one or more candidate ASes. 
Repeated linking yields disjoint clusters, each representing a unified anomaly 
detailing the affected prefixes, responsible ASes, and temporal bounds. 
This methodology yields concise, AS-attributed anomaly records, supporting targeted investigation and remediation.

\section{Experimental Results}
\label{sec:experiment}


In this section, we conduct comprehensive evaluations on real-world BGP data to verify the effectiveness of BGPShield, which comprises two modules, \emph{i.e.,} ALBE and BAD. We first analyze whether the representations generated by ALBE capture the two semantic features introduced in Section~\ref{sec:embedding-gen}.
We then evaluate the effectiveness of BAD by examining the SAM-ED algorithm in identifying anomalous route changes. Finally, we assess the overall performance of BGPShield against SOTA and advanced methods, along with its generalizability and computational overhead.

\subsection{Analysis of Representations}
\label{sec:embed-analysis}

\begin{figure}[t!]
    \centering
    \includegraphics[width=\linewidth]{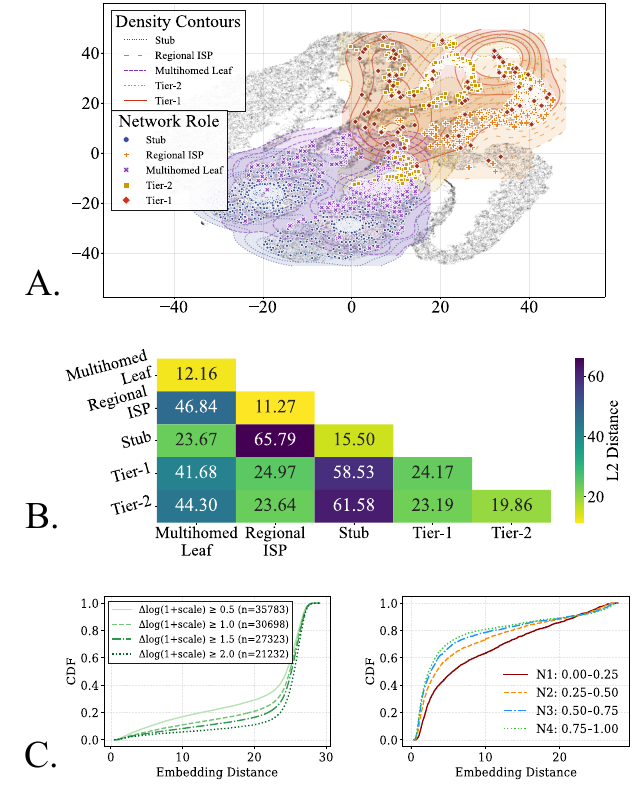}
    \caption{\textbf{Analysis of AS Embedding Semantics.}
    \textbf{(A)} Visualization of AS Roles with density contours highlighting their distribution. \textbf{(B)} Heatmap of pairwise L2 distances between AS Roles; diagonal and off-diagonal entries respectively represent intra- and inter-type distances, where smaller distances indicate higher similarity. \textbf{(C)} Embedding distance CDFs grouped by: (left) operational scale difference and (right) neighbor overlap coefficient.}
    \label{fig:dist-metric}
\end{figure}

\begin{figure*}[!t]
    \centering
    \includegraphics[width=1\textwidth, height=0.35\textheight]{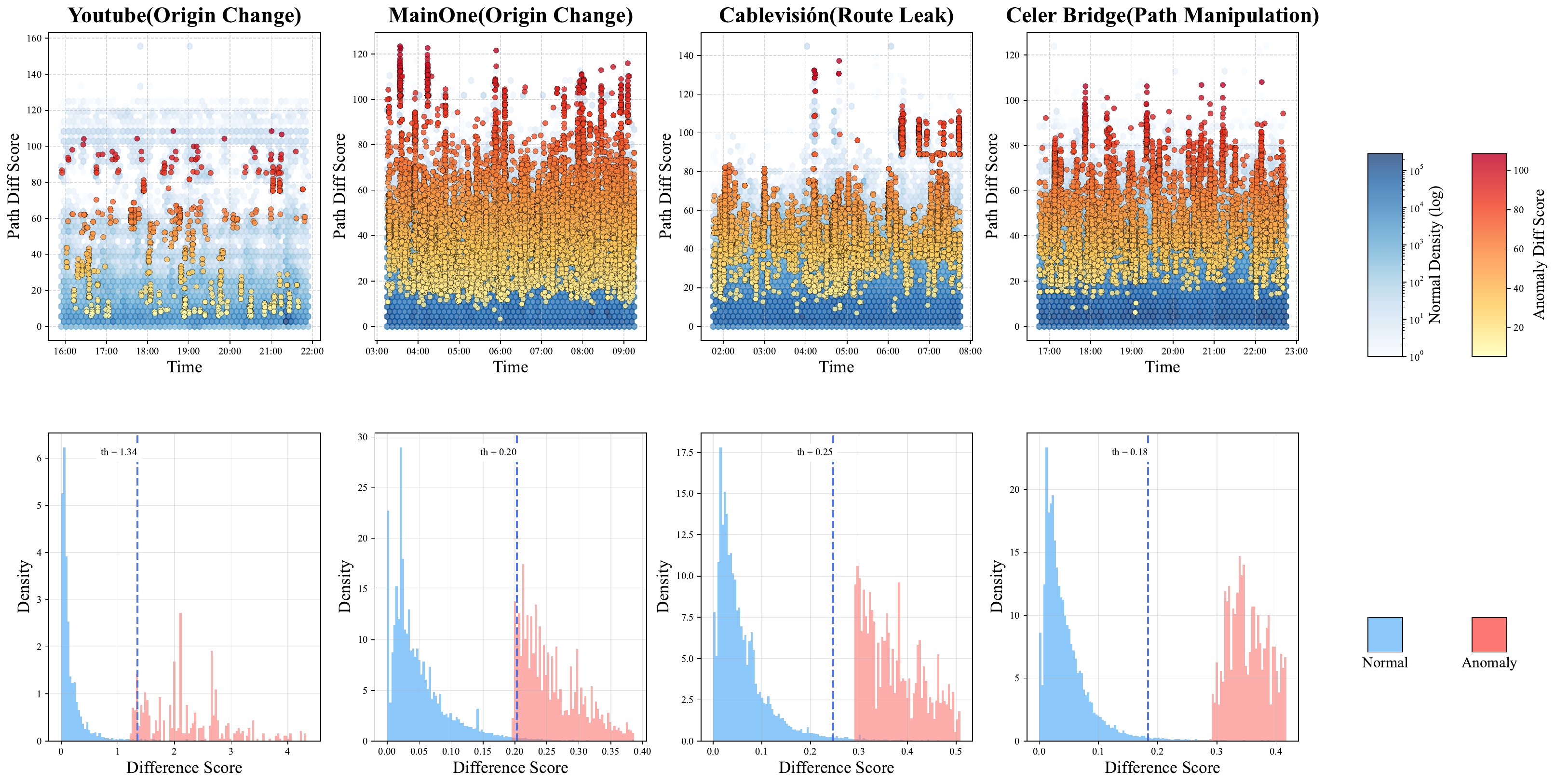}
    \caption{\textbf{Distribution Characteristics of Path Difference Scores Across Four Randomly Selected Datasets}}
    \label{fig:path_diff_dynamic_threshold}
    \vspace{-0.5cm}
\end{figure*}

The effectiveness of ALBE is measured by how well the generated embeddings encode and distinguish the behavioral semantics (mentioned in Section~\ref{sec:intro}) of ASes.
To rigorously evaluate the behavioral semantics, we treat the learned features as conceptual attributes and examine their consistency with the spatial distribution of the latent space.

We first analyze \textbf{Behavioral Portrait} (Feature A), which posits that embeddings should categorize ASes based on their operational behaviors (\emph{e.g.,} operational scale) rather than just topology structural patterns or statistical regularities. 
To establish a reliable ground truth for this validation, we construct a Gold Standard subset of 2,190 ASes where operational roles are confirmed through the cross validation of authorized platforms (\emph{e.g.,} PeeringDB and CAIDA). 
This subset (comprising 90 Tier-1, 300 Tier-2, 600 Regional ISP, 500 Multihomed Leaf, and 700 Stub networks) ensures that the evaluation is grounded in ASes with unambiguous functional roles. The resulting embeddings are visualized using t-SNE~\cite{van2008visualizing} in Figure~\ref{fig:dist-metric}(A).
Figure~\ref{fig:dist-metric}(A\&B) suggest that the embedding distances within the latent BGP space align with the operational logic of the Internet. 
Tier-1 and Tier-2 providers form adjacent, overlapping clusters with a low inter-type distance of 23.19, which appears consistent with their shared roles as global transit providers. 
In contrast, Stub ASes are embedded far from the routing core, yielding a distance of 58.53 from Tier-1 networks. 
A prominent observation is the distance of 65.79 between Regional ISPs and Stubs. 
Despite having comparable connectivity levels in the AS graph, these two roles exhibit fundamentally different routing behaviors (Regional ISPs operate as transit providers that must actively export routes between their customers and providers while Stubs function as leaf nodes that typically only announce their own prefixes and refuse to carry third party traffic). This significant separation suggests that the latent space is sensitive to functional roles that extend beyond mere topological structures.
Further population-level evidence is provided in Figure~\ref{fig:dist-metric}(C), which reveals how operational scale and structural similarity influence the latent space. Figure~\ref{fig:dist-metric}(C) left shows that embedding distance grows with operational scale disparity, indicating that LSE is responsive to scale as a semantic attribute. 
Moreover, Figure~\ref{fig:dist-metric}(C) right shows that higher neighbor overlap naturally leads to smaller distances. 
However, the marginal separation between consecutive overlap groups appears to diminish as structural similarity increases. Specifically, the shift from N1 to N2 is notably larger than the shift from N3 to N4. This diminishing effect suggests that topological overlap may have a bounded influence on the latent representation. Beyond a certain structural threshold, non-topological factors such as operational mandates and global roles likely dominate the remaining distance. 
Finally, we consider \textbf{Routing Policy Rationale} (Feature B), which represents stable policy rationales (\emph{e.g.,} the underlying policy rationales underpinning routing decisions). As detailed in Section~\ref{sec:realtime_detection}, the embeddings sustain detection effectiveness over five years of network evolution. Such persistence would be highly improbable if the model captured only transient topological features, whereas the stability of policy level logic provides a plausible explanation for this persistent accuracy.

\subsection{Evaluation of Path Difference Score}
\label{sec:path-diff-evaluation}

The effectiveness of BGP Anomaly Detector (BAD) mainly depends on the capacity of the SAM-ED algorithm to accurately measure path difference scores.
To evaluate this algorithm, we conduct experiments on the real-world BGP anomaly datasets introduced in Section~\ref{sec:dataset}  

Figure~\ref{fig:path_diff_dynamic_threshold} demonstrates the characteristics of path difference score distributions across four randomly selected datasets. From a temporal perspective, the scores associated with benign routing changes follow approximately uniform distribution throughout the observation period, whereas those linked to anomalies show distinct clustering with high-magnitude outliers concentrated around specific timestamps. This phenomenon reflects the characteristic signature of historical anomaly events where a single incident typically triggers simultaneous changes across multiple routing paths, leading to clusters of elevated path difference scores within narrow time windows. The density distributions and dynamic threshold illustrated in Figure~\ref{fig:path_diff_dynamic_threshold} further highlight that the path difference scores of anomalous routing changes  significantly exceed those of legitimate changes, 
revealing the substantial disruption that routing anomalies cause to AS role assignments.
In contrast, scores from legitimate routing changes are significantly lower and follow a distribution resembling a Gaussian profile, precisely because these route changes introduce only limited structural variation to the path. 
Generally, legitimate routing changes typically occur between ASes with comparable preferences in routing behaviors and are driven by controllable and localized operational adjustments, which tend to preserve the semantic consistency of the origin AS path. Thus, they introduce minimal perturbation to the distribution of routing roles along the path and do not trigger significant deviations in path difference metrics.

BAD effectively separates benign and anomalous routing changes with a dynamic threshold. The dynamic threshold employs $\mu + 4 \sigma$ rule, covering over 99.7\% of normal data. 
As shown in Figure~\ref{fig:path_diff_dynamic_threshold}, the dynamic threshold precisely demarcates anomalous updates from legitimate ones, thereby establishing a reliable foundation for anomaly detection.

\subsection{Overall Performance of BGPShield}
\label{sec:overall-performance}

\begin{table*}[!t]\centering
\renewcommand{\arraystretch}{1.15}

\caption{\textbf{Detection Results on 16 Real-world Datasets.} Values denote \emph{True Positive Alarms} (\emph{False Positive Alarms}); \(\ddag\) marks datasets whose confirmed anomalies remain \emph{undetected}. Rows with \colorbox{blue!20}{blue background} indicate incidents occurred after the latest public release of the LLMs employed by BGPShield.}
\label{tab:anomaly_detection_results}

\resizebox{\textwidth}{!}{%
\begin{tabular}{lllrrrrrrr}
\toprule
\textbf{Dataset} & \textbf{Date} & \textbf{Prefix} & \textbf{ARES} & \textbf{AP2VEC} & \textbf{BEAM} & \textbf{Q1.7B} & \textbf{Q1.7B} & \textbf{DS-8B} & \textbf{DS-8B} \\
& \textbf{(mm/dd/yy)}& & & & & & \textbf{(A)} & & \textbf{(A)}\\
\midrule
\multicolumn{10}{@{}l@{}}{\small\textit{\textbf{\textcolor{gray}{Origin Change}}}} \\
YouTube & 02/24/08 & 208.65.153.0/24 & 54(0)\textsuperscript{\ddag} & 664(182) & 620(32) & 1324(100) & 1366(208) & 1225(172) & 641(28) \\
Petersburg & 01/07/15 & 192.135.33.0/24 & 166(0)\textsuperscript{\ddag} & 1842(666) & 1868(118) & 3272(2) & 2738(1) & 3582(4) & 2423(3) \\
Rostelecom & 04/01/20 & 104.18.216.0/21 & 640(4)\textsuperscript{\ddag} & 1896(586)\textsuperscript{\ddag} & 1904(105) & 2671(44) & 2671(28) & 2206(26) & 2148(8) \\
BackConn\_3 & 05/20/16 & 191.86.129.0/24 & 1230(2)\textsuperscript{\ddag} & 1750(722)\textsuperscript{\ddag} & 2692(22) & 1453(6) & 1512(4) & 1412(11) & 2795(2) \\
MainOne & 07/30/18 & 91.108.4.0/24 & 449(1)\textsuperscript{\ddag} & 2129(505)\textsuperscript{\ddag} & 2555(23) & 4374(131)\textsuperscript{\ddag} & 3644(33) & 3135(50) & 4601(6) \\
\rowcolor{blue!20}
CTRLS & 11/06/25 & 5.160.214.0/24 & 1063(0)\textsuperscript{\ddag} & 576(192)\textsuperscript{\ddag} & 171(4) & 593(2) & 820(8) & 673(2) & 1070(8) \\
\multicolumn{10}{@{}l@{}}{\small\textit{\textbf{\textcolor{gray}{Path Manipulation}}}} \\
Bitcanal & 01/07/15 & 115.116.96.0/24 & 70(0)\textsuperscript{\ddag} & 1842(666) & 2101(139) & 3397(2) & 2875(1) & 3739(4) & 2866(1) \\
CelerBridge & 08/17/22 & 44.235.216.0/24 & 481(0) & 2097(1138)\textsuperscript{\ddag} & 2414(51)\textsuperscript{\ddag} & 2697(8) & 3078(2) & 2831(0) & 2694(0) \\
DEFCON & 08/10/08 & 24.120.56,58.0/24 & 37(0) & 491(216) & 816(348) & 1628(372) & 590(113) & 1120(298) & 494(22) \\
BackConn\_1 & 02/20/16 & 72.20.0.0/24 & 166(2)\textsuperscript{\ddag} & 2209(792) & 2565(39) & 2084(24) & 2397(23) & 2198(32) & 2923(9) \\
BackConn\_2 & 04/16/16 & 161.123.172.0/24 & 616(1) & 1378(450) & 3027(32) & 2781(116) & 1999(45) & 1860(40) & 3291(5) \\
\multicolumn{10}{@{}l@{}}{\small\textit{\textbf{\textcolor{gray}{Route Leak}}}} \\
Malaysia & 06/12/15 & 31.13.67.0/24 & 5256(118) & 1463(588)\textsuperscript{\ddag} & 1818(121) & 1959(2) & 1627(3) & 1564(2) & 2382(3) \\
Cablevisión & 02/11/21 & 201.157.24.0/24 & 953(11)\textsuperscript{\ddag} & 2088(712)\textsuperscript{\ddag} & 2314(51) & 2697(8) & 3078(2) & 2831(0) & 2594(0) \\
Vodafone & 04/16/21 & 24.152.117.0/24 & 866(1) & 3855(1682)\textsuperscript{\ddag} & 2670(40)\textsuperscript{\ddag} & 2891(18)\textsuperscript{\ddag} & 3493(13) & 2916(22)\textsuperscript{\ddag} & 3716(2) \\
\rowcolor{blue!20}
SINGTEL & 09/16/25 & 89.185.86.0/23 & 1421(0)\textsuperscript{\ddag} & 1019(627)\textsuperscript{\ddag} & 162(2)\textsuperscript{\ddag} & 484(1) & 637(4) & 613(2) & 1081(5) \\
\rowcolor{blue!20}
Loviz & 10/10/25 & 1.193.210.0/24 & 1067(1)\textsuperscript{\ddag} & 782(335) & 133(1)\textsuperscript{\ddag} & 464(4) & 742(4) & 520(1) & 881(5) \\
\midrule
\textbf{Detected} & & & \textbf{5/16} & \textbf{7/16} & \textbf{12/16} & \textbf{14/16} & \textbf{16/16} & \textbf{15/16} & \textbf{16/16} \\
\bottomrule
\end{tabular}%
}
\end{table*}


We validate the performance of the BGPShield system with particular emphasis on evaluating its effectiveness in identifying realistic anomalies. Specifically, we conduct the experiments on 16 datasets (detailed in Section~\ref{sec:path-diff-evaluation}), covering various BGP anomaly types including Origin Change, Path Manipulation, and Route Leak, with each dataset corresponding to publicly reported BGP incidents. 



For a comprehensive performance evaluation, we instantiate the semantic encoding module of ALBE two advanced generative language models and conduct an ablation study to isolate the impact of the BGP Adapter:

\begin{itemize}[itemsep=0em, left=0pt]
    \item \textbf{DS-8B}. Generating high-dimensional embeddings using DeepSeek-R1-0528-Qwen3-8B;
    \item \textbf{DS-8B (A)}. The fully equipped BGPShield variant, which applies the BGP Adapter to adapt DS-8B embeddings into BGP domain space.
    \item \textbf{Q1.7B}. Generating embeddings using Qwen3-1.7B;
    \item \textbf{Q1.7B (A)}. Applying the Adapter to Q1.7B embeddings.
\end{itemize}

Notably, we evaluate three advanced anomaly detection methods: ARES (2025)~\cite{ares}, AP2VEC (2022)~\cite{9754706} and BEAM (2024, SOTA)~\cite{298068}. 
To ensure an unbiased assessment, each of these methods is evaluated using the default parameter configurations specified in its original implementation.
For a fair comparison, we integrate these methods and BGPShield into a unified anomaly attribution framework. This framework consolidates detection results into alarm-level events and identifies the responsible ASes based on standardized criteria for anomaly attribution.
We then apply a post-processing pipeline to ensure consistency in how anomalies are grouped, and evaluated across different systems (detailed in Section~\ref{sec:event-aggregation}).
We do not include active probing based methods in our comparison as they require live network access and contemporaneous measurements, making them inherently unsuitable for retrospective analysis of historical datasets.

Note that each detection system generates multiple alerts within each evaluation dataset, and a single alert can serve as an aggregate of all BGP updates associated with a potential routing anomaly. Thus, we consider an alert valid if it matches confirmed information, particularly when the system identifies the target prefix as anomalous and accurately pinpoints the misbehaving AS as the responsible party, indicating successful detection of the confirmed anomaly for that event.  
Beyond confirmed anomalies, other alerts may indicate previously unrevealed anomaly incidents or represent false positives. To validate these unconfirmed incidents, we extend the minor anomaly identification methodology introduced in BEAM~\cite{298068} and redefine five typical anomalous routing change patterns for matching against suspicious routing behaviors (these patterns have been endorsed by cybersecurity experts):


\begin{itemize}[itemsep=0em, left=0pt]
    \item Origin Change: routing changes where origin ASes before and after the change belong to different organizations with different RPKI validation statuses;
    \item Route Leaks: routing paths before or after changes violate valley-free principles;
    \item Path Manipulation: routing paths contain reserved ASNs or adjacent ASes without commercial relationships;
    \item ROA Misconfiguration: origin ASes belong to the same organization but exhibit different RPKI validation statuses;
    \item Weak Path Tampering: paths show tampering suspicions but are classified as potential anomalies (such as abnormal AS insertion) due to unqueryable RPKI status.
\end{itemize}



If an anomalous event can be categorized as any of these five patterns, it is classified as a True Positive (TP); otherwise, it is considered a False Positive (FP).
As illustrated in Table~\ref{tab:anomaly_detection_results}, the adapter-equipped BGPShield variants (DS-8B (A) and Q1.7B (A)) successfully detect all 16 confirmed routing anomalies with a low FDR. Among these, three incidents were unseen during the pre-training process of our employed LLMs, and all positive alarms related to these events were carefully verified by cybersecurity experts. Furthermore, \textbf{the ablation study explicitly highlights the effectiveness of the BGP Adapter}: raw LLM embeddings (DS-8B and Q1.7B) fail to capture all incidents (missing 1 and 2 confirmed anomalies, respectively) and yield a substantially higher FDR, whereas  applying the BGP Adapter achieves 100\% detection with significantly suppressed false positives.
Notably, even with a relatively compact 1.7B parameter model, BGPShield (Q1.7B (A)) still achieves exceptional performance, demonstrating that ALBE enables effective anomaly detection without requiring massive computational resources. This efficiency advantage makes our solution practically deployable in resource-constrained environments while maintaining SOTA detection precision.
Regarding Ares, despite its reported strong performance, its detection parameters are likely optimized for its own curated datasets, which can explain the relatively weaker performance observed on our diverse real-world datasets.
In summary, although other evaluated methods have competitive performance in several cases, BGPShield achieves more comprehensive anomaly coverage with lower FDR.

\subsection{Real-Time Detection Evaluation}
\label{sec:realtime_detection}


\begin{figure}[t!]
\centering
\includegraphics[width=\linewidth]{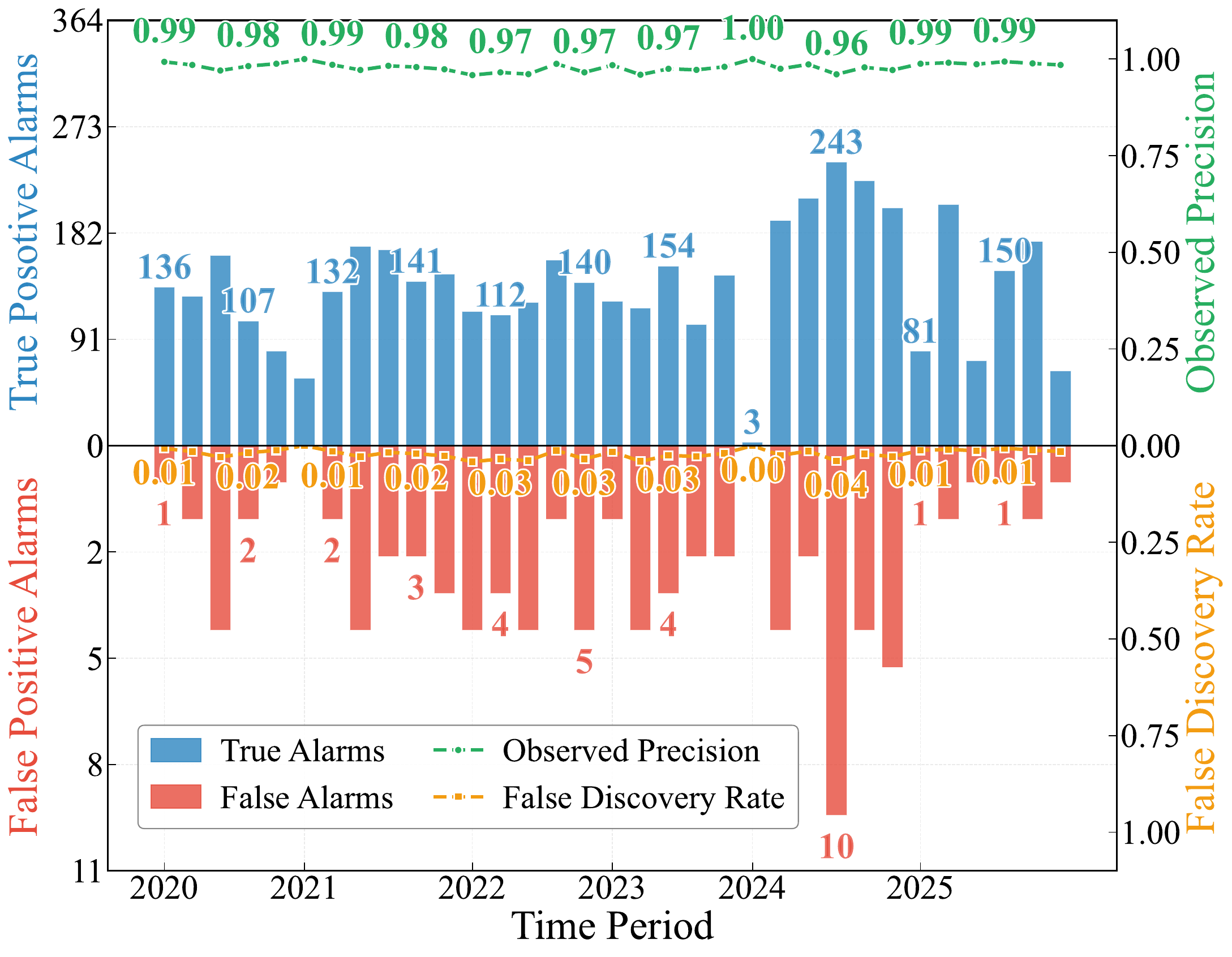}
\caption{\textbf{Generalization performance over time.} (Top) true alarms and precision; (bottom) false alarms and FDR.}
\label{fig:genralization}
\end{figure}


\textbf{Runtime Overhead.} 
To further assess BGPShield's real-time processing capabilities, we conduct comprehensive performance tests on a Linux server equipped with an Intel\textsuperscript{\textregistered} Xeon\textsuperscript{\textregistered} Gold 6226R CPU@2.90GHz. The experiments utilize a real-world dataset comprising 1,574,063 routing records collected from the \textit{WIDE} collector over a 24-hour period on October 12, 2024. 
BGPShield demonstrates significant computational advantages over the fastest SOTA approach (BEAM). In terms of cold-start overhead (generating representations from scratch), BEAM requires 71 hours to construct all representations, while BGPShield completes the same task in 31 hours. 
In terms of real-time adaptation, BEAM necessities a thorough retraining of the entire AS graph whenever it encounters a previously unseen AS, requiring 65h on average. 
In contrast, BGPShield can generate embeddings for each AS independently. This capability allows BGPShield to construct a new representation in less than one second (0.6 seconds on average). demonstrating its outstanding adaptability to evolving network conditions.

\noindent\textbf{Generalization Capability.} To validate the generalizability of BGPShield in dynamic BGP environments, we evaluate routing events from randomly selected months spanning from 2021 to 2025 using 16-dimensional embeddings trained on 2020 data. 
For newly observed ASes, we independently generating representations 
for each AS without reprocessing existing ASes.
As illustrated in Figure~\ref{fig:genralization}, the system maintains an average precision of 97.94\% with a minimum precision of 95.83\%. The FDR remains consistently below 4.20\%, averaging 2.06\% (even 0 in some periods). Overall, BGPShield exhibits excellent temporal generalizability with minimal performance degradation over the five-year evaluation period, significantly outperforming SOTA methods. 
Notably, BGPShield actually has better capabilities than what the experimental results as many apparent false positives arise from missing or delayed RPKI validation, reflecting limitations of the RPKI ecosystem rather than detector errors.
Temporal gaps between ROA publication and ROV deployment, along with legacy configuration issues that mark valid routes as invalid, cause some legitimately anomalous or normal routes to appear unvalidated.
\looseness=-1

\subsection{Robustness Analysis}
\label{sec:robust-analysis}

We evaluate the robustness of BGPShield against noisy AS relationship data by introducing perturbations into the original AS Relationship dataset, then regenerating representations on each perturbed dataset, and evaluating their performance on the datasets introduced in Section~\ref{sec:dataset}. This evaluation specifically tests whether the embeddings can maintain detection precision under corrupted information.

\begin{figure}[t]
\centering
\includegraphics[width=\linewidth]{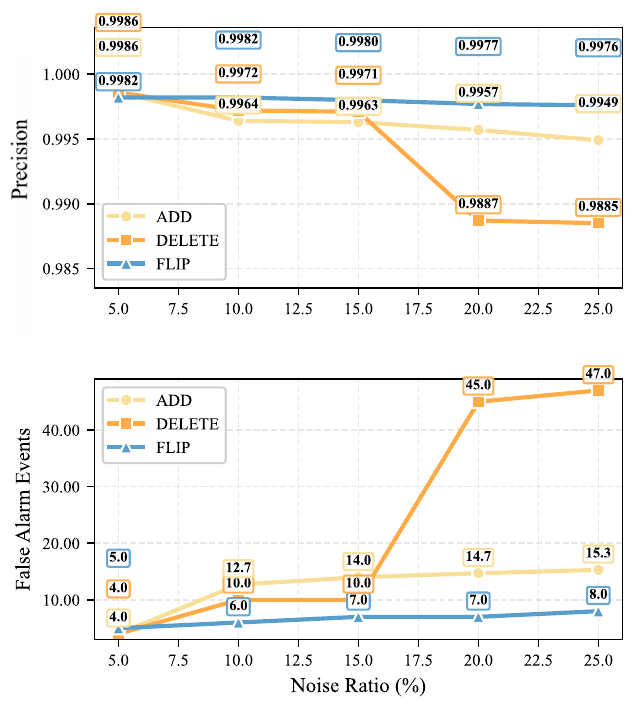}
 \caption{\textbf{Robustness Evaluation Results.} \emph{Precision} degradation (top) and \emph{False Alarm Events} degradation (bottom) across varying noise conditions.}
 \label{fig:robustness}
 \vspace{-0.5cm}
\end{figure}

After generating representations on the perturbed data, we conduct all evaluations with the same steps described in Section~\ref{sec:experiment}. To avoid bias and ensure statistical reliability, we repeat each experiment three times and report the averaged results.
As illustrated in Figure~\ref{fig:robustness}, experimental results confirm that BGPShield remains highly robust to substantial perturbations. Specifically, BGPShield maintains over 99\% precision even when the noise ratio reaches 15\%, indicating strong resilience to moderate data corruption. Even under extreme noise conditions with 25\% perturbation (a scenario unlikely to occur in practice), BGPShield still sustains 98.8\% precision while generating fewer than 50 false alarms.
The exceptional robustness mainly stems from BGPShield's ability to encode underlying \textbf{routing policy rationale} (Feature B) beyond mere behavioral features. 
When topology is perturbed (\emph{e.g.,} relationships deleted or flipped), the embeddings can still infer correct routing tendencies based on operational semantics and policy patterns.
\looseness=-1

Performance varies slightly across noise types, with deletion perturbations resulting in approximately 1.0\% accuracy reduction at 25\% noise ratio, while addition and flip perturbations show even smaller reductions of 0.4\% and 0.6\% respectively. This pattern reflects that while complete structural information is ideal, the semantic richness of our embeddings provides substantial redundancy against missing or incorrect topological data.
The consistent high performance under extreme noise conditions demonstrates the practical superiority of BGPShield. Real-world AS relationship inference typically exhibits much lower error rates than our tested extreme conditions, suggesting that BGPShield's semantic embeddings can maintain effect in evolving operational environments over the long-term.

\section{Case Study}
\label{sec:case}

\begin{figure}[t]
\centering
\includegraphics[width=\linewidth]{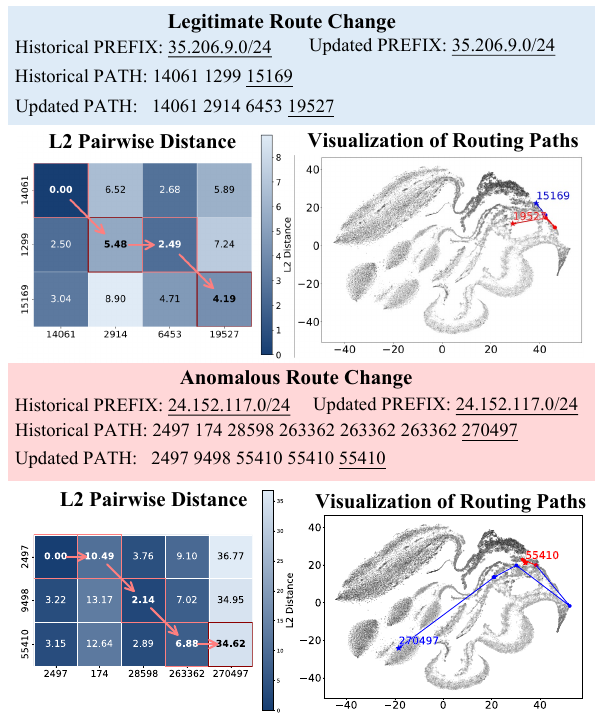}
\caption{\label{fig:case}\textbf{Comparison of Legitimate and Anomalous BGP Route Changes Detected by BAD.} \emph{L2 Pairwise Distance} d(left) show similarity between AS pairs (smaller distances indicate higher similarity), while \emph{path visualizations} (right) display routing trajectories.}
\vspace{-0.5cm}
\end{figure}

\textbf{Case Background and Data Preparation}. On April 16, 2021, a major BGP route leak incident occurred~\cite{major2021}, originating from Vodafone India (AS55410). This event involved the unauthorized propagation of the prefix 24.152.117.0/24 route, which was originally announced by AS270497 (RUTE MARIA DA CUNHA). The anomalous update shortened the AS path and diverted traffic, potentially affecting global routing stability.
In this case, the key ASes include:
\begin{itemize}[itemsep=0em, align=parleft, left=0pt]
    \item AS270497 (RUTE MARIA DA CUNHA): The legitimate origin AS claiming ownership of prefix 24.152.117.0/24
    \item AS55410 (Vodafone Idea Ltd.): The leak source, becoming the new origin AS.
\end{itemize}

To analyze this incident, BGPShield first collects all BGP metadata associated with ASes. ALBE (detailed in Section~\ref{sec:embedding-gen}) then formulates structured AS profiles from the heterogeneous metadata through the segment-wise aggregation scheme with instruction prompts, and feeds them into a pre-trained LLM to parse routing-relevant information in natural language form. The BGP Adapter subsequently adapts the LLM embeddings into the BGP domain space, generating semantic embeddings that reflect the Behavioral Portrait and Routing Policy Rationale of each AS.

\noindent\textbf{Anomaly Detection}. 
Based on the embeddings, BAD further processes BGP updates for prefix 24.152.117.0/24, where the update changed the path from \textit{Historical path:} {{2497 174 28598 263362 263362 263362 270497}
to \textit{Updated path:} {2497 9498 55410 55410 55410}. \looseness=-1

\textit{Step 1} (Path Difference Scoring).
For this case, BAD first retrieves the semantic embeddings produced by ALBE for each AS in both the original and updated 
paths. SAM-ED then flexibly aligns the two embedding sequences and accumulates the pairwise distances between aligned AS embeddings as the path difference score, capturing the cumulative semantic deviation between the two paths in terms of both behavioral and policy-level characteristics. The arrows in the heatmap of Figure ~\ref{fig:case} indicate the aligned path identified by SAM-ED.

\textit{Step 2} (Anomalous Route Detection). 
For this case, BAD compares its path difference score against the dynamic threshold (detailed in Section~\ref{sec:suspicious-detection}). If this score exceeds the dynamic threshold, it immediately triggers an anomalous route change classification. 
the red path (anomalous) exhibits a geometric distribution distinctly different from historical path patterns, with significant spatial deviation from the blue path (normal). The L2 Pairwise Distance heatmap in Figure ~\ref{fig:case} quantitatively captures this deviation, reporting a distance of 34.62 for the semantically dissimilar AS pair (AS55410 \& AS270497), in contrast to a substantially lower distance (4.19) observed for the semantically similar AS pair (AS15169 \& AS19527) in legitimate route changes. This contrast directly reflects the effectiveness of ALBE to place behaviorally distinct ASes far apart and functionally similar ASes close together in the BGP domain space.


\textit{Step 3} (Multi-perspective Aggregation). 
Once an anomalous route change classification is triggered, BAD aggregates related observations from multiple vantage points to confirm and localize the event. 
By grouping these related anomalous route changes, BAD correctly identifies them as belonging to a single, distinct anomalous event.

\section{Conclusion}                                                                                
\label{sec:conclusion}

In this paper, we propose BGPShield, a novel anomaly detection framework built on semantic representations in the BGP domain space that capture both the \emph{Behavioral Portrait} and \emph{Routing Policy Rationale} of each AS. Specifically, we propose an Adaptive LLM BGP Encoder (ALBE) trained on a contrastive loss to embed AS profiles into the
BGP embeddings. 
ALBE captures the routing behaviors and routing semantics beyond topology, supporting embedding generation for a previously unseen AS within one second without retraining.
Based on ALBE's representations, we introduce the SAM-ED algorithm that quantifies semantic deviations between historical and updated
paths through two BGP-aware designs, enabling BGPShield to accurately and reliably distinguish anomalies from routine variations. 

We further implement a prototype of BGPShield and evaluate its performance on 16 real-world RouteViews Datasets. The results show that BGPShield achieves 100\% detection of verified anomalies with an average FDR 2–3$\times$ lower than the SOTA method and maintains 98.8\% precision under 25\% AS-level information noise, demonstrating high detection precision and robust generalizability to evolving networks and unseen events.
Moreover, BGPShield can construct the representation of a previously unseen AS within one second, significantly outperforming SOTA BEAM that requires costly retraining on the entire AS graph.

\bibliographystyle{IEEEtran}                                                                        
\bibliography{reference}
   
\end{document}